\documentclass{emulateapj}
\usepackage{amsmath,fleqn,epsfig}

\newcommand {\kms}{\,{\rm km}\,{\rm s}^{-1}}

\newcommand {\masyr}{\,{\rm mas\,yr^{-1}}}
\newcommand {\magn}{\,{\rm mag}}

\newcommand {\Gyr}{\,{\rm Gyr}}

\newcommand {\kpc}{\,{\rm kpc}}

\newcommand {\V}{{\rm V}}

\newcommand {\dex}{\,{\rm dex}}
\newcommand {\llg}{\log {\rm g}}

\newcommand {\Rsun}{{R_{G,\odot}}}
\newcommand {\feh}{\hbox{[Fe/H]}}
\newcommand {\feha}{{\hbox{[Fe/H]}}_{\rm DR7}}

\newcommand{\beq}{\begin{equation}}
\newcommand{\eeq}{\end{equation}}

\shorttitle{Analysing halo duality}
\shortauthors{Sch\"onrich, Asplund \& Casagrande}

\begin{document}

\title{Does SEGUE/SDSS indicate a dual Galactic halo?}

\author{Ralph Sch\"onrich\altaffilmark{1,2}}
\affil{Department of Astronomy, The Ohio State University, Columbus, OH 43210}
\affil{Rudolf Peierls Centre for Theoretical Physics, University of Oxford, Oxford, UK}
\email{Email: ralph.schoenrich@physics.ox.ac.uk}

\author{Martin Asplund\altaffilmark{3} and Luca Casagrande\altaffilmark{3}}
\affil{Research School of Astronomy \& Astrophysics, Australian National University, Canberra, Australia}

\altaffiltext{1}{Hubble Fellow, Department of Astronomy, The Ohio State University, 140 West 18th Avenue, Columbus, OH 43210-1173, USA}
\altaffiltext{2}{Rudolf Peierls Centre for Theoretical Physics, 1 Keble Road, Oxford OX1 3NP, UK}
\altaffiltext{3}{Research School of Astronomy \& Astrophysics, Australian National University, Cotter Road, Weston, ACT 2611, Australia}


\begin{abstract}

We re-examine recent claims of observational evidence for a dual Galactic halo in SEGUE/SDSS data, and trace them back to improper error treatment and neglect of selection effects. In particular, the detection of a vertical abundance gradient in the halo can be explained as a metallicity bias in distance. A similar bias, and the impact of disk contamination, affect the sample of blue horizontal branch stars. These examples highlight why non-volume complete samples require forward-modelling from theoretical models or extensive bias-corrections.
We also show how observational uncertainties produce the specific non-Gaussianity in the observed
azimuthal velocity distribution of halo stars, which can be erroneously identified as two Gaussian components. A single kinematic component yields an excellent fit to the observed data, when we model the measurement process including distance uncertainties.
Furthermore, we show that sample differences in proper motion space are the direct consequence of kinematic cuts, and are enhanced when distance estimates are less accurate. Thus, their presence is neither a proof for a separate population, nor a measure of reliability for the applied distances.
We conclude that currently there is no evidence from SEGUE/SDSS that would favour a dual Galactic halo over a single halo full of substructure.

\end{abstract}


\section{Introduction}
Since the discovery of the Galactic halo \citep[][]{Schwarzschild52, Eggen62}, its structure and origin have been under intense debate. Three main sources of halo stars have been discussed in the past \citep[see][ for a discussion]{Sheffield12}: monolithic collapse in early galaxy formation as put forth by \cite{Eggen62};
the accretion of satellites \citep[][]{Searle78}; and kick-up of disc/bulge stars via minor mergers \citep[e.g.][]{Purcell10}, ejections from cluster cores \citep[e.g.][]{Leonard91}, or binary interactions \citep[][]{Przybilla08} involving a supermassive or intermediate mass black hole \citep[e.g.][]{Hills88, Gualandris07}. 
Of these the second is considered to be the dominant contribution to the Galactic halo based on $\Lambda CDM$ cosmological simulations and from observational data 
\citep[see][]{Sheffield12}, The impact of kicked out stars is strongly limited theoretically, as well as observationally by the low number of metal-rich halo stars. 

It can be hypothesized that stars from an initial collapse should on average carry some prograde momentum. For accreted stars, early studies \citep[][]{Quinn86, Byrd86}, recently confirmed by \citep[][]{Murante10}, suggested that dynamic friction differentiates between prograde and retrograde infalling satellites, leaving behind a potentially detectable asymmetry in kinematics. It is, however, not clear, how this signature should translate into different rotation as a function of metallicity. \cite[][]{Cooper10} found that the accretion signatures along with radial abundance gradients vary strongly in simulated galaxies, depending strongly on the individual accretion history. 

In principle three different observational signatures have been proposed:
\begin{itemize} 
\item Asymmetry in the (azimuthal) velocity distribution, which may be a consequence of dynamic friction. However, other processes as well as the special accretion history can lead to such a distribution. This dates back to \cite[][]{Norris89}, but it has been known since \cite{Strom27} and later \cite{Ryan92} that distance errors can account for this type of asymmetries.
\item Radial metallicity gradients. As noted before, these depend on the specific accretion history of a Galaxy. By tendency the material accreted later onto a galaxy stems from less dense regions that evolved later and produced less massive stellar systems \citep[this was already pointed out by][]{Kant1755} and hence are likely to have a lower metallicity. Claims of related differences between the inner and outer halo were made as early as \cite{Searle78} and \cite{Preston90}.
\item Radial gradients in angular momentum/ mean azimuthal velocity. Again, this hinges on the detailed accretion history, correlating velocities and metallicities to varying degree and sign. For our Galaxy, BHB stars set quite narrow limits on any radial trend \citep[][]{FSII}.
\end{itemize}

\cite{Carollo07} and \cite{Carollo10} (hereafter C07 and C10) claimed the existence of a dual halo that consists of a prograde, metal-rich ($\feh \approx -1.6$), inner halo and a distinct retrograde, more metal-poor ($\feh \approx -2.2$), outer halo. This observational result triggered a major series of theoretical papers finding such structures in simulations \citep[e.g.][]{Zolotov09, Zolotov10, Font11, McCarthy12, Tissera12}, while some other studies \citep[e.g.][]{Lucia08, Cooper10} find no convincing trends. 
The result was criticized by \cite{SAC} (hereafter SAC11),tracing their claims back to the inappropriate use of Gaussian analysis and the neglect of observational errors, as well as unphysical distance estimates in their sample. Accounting for these effects, SAC11 showed that on any reasonable adopted distance scale, including C10's own distances, their findings of a halo duality vanished. Recently \cite{Beers12} (hereafter B12) published a re-analysis affirmative of the original C07,10 studies, finding again a dual halo with a retrograde metal-poor component.

It could be argued that this question will soon be answered by the Gaia satellite mission. While this is likely true, the purpose of the present paper is more general, i.e. to lay out on the example of B12 the many subtleties due to selection biases and error analysis, which all studies of this kind even with Gaia data need to address in order to derive meaningful conclusions. With this spirit we make the data used in this work publicly available\footnote{Please find the data under \newline {\it \scriptsize http://www-thphys.physics.ox.ac.uk/people/RalphSchoenrich/data/halo/}. We will be delighted to provide any additional information upon request.} to enable independent investigations on the topic by others.

In this work we will focus on the question of halo duality and evaluate the evidence presented so far. We will focus on the arguments as laid out by B12 and rely on the distance scale of B12, since in SAC11 we already did an extensive study of other available distance prescriptions. In Section \ref{sec:replica} we describe our attempt to replicate those distance assignments and discuss the distance calibrations of \cite{Beers00}. We examine the central points of observational evidence as put forth by B12 in Section \ref{sec:arguments}. Further arguments that have been used in favour of a dual halo detection are evaluated in Section \ref{sec:evidence}. In Section \ref{sec:conclude} we summarize our conclusions: none of the arguments put forward by C07, C10 and B12 hold up to closer scrutiny, and therefore, at present, there is no evidence for a dual halo structure with a retrograde metal-poor outer halo component in the SEGUE/SDSS data.

\section{Replication of the Beers et al. sample}\label{sec:replica}

As in the previous studies we use the sample of photometric and reddening standard stars in SEGUE \citep[][]{Yanny09}. The sample selection is identical to SAC11; in addition we remove stars with expected reddening of more than $0.75 \magn$ in the $g$-band and apply the proper motion and radial velocity corrections given by \cite{S12} - despite being necessary for obtaining a clean sample, neither of these steps has a significant impact on the analysis.

In SAC11 we did an extensive study of dual halo signatures on four different distance scales bracketing the reasonable range of scales in the literature, and found no hint of duality in any of them. Since B12 used the difference in distance calibrations to argue why their finding of a dual halo is still correct, we will now solely examine their evidence on their own distance scale.  We note that the one distance scale that B12 wrongly ascribed to us and made a central element of their criticism, is in fact not ours: it is $\sim 10 \%$ shorter than the adaptation of the \cite{Ivz08} distance calibration we used as one of four distance scales (see the Appendix for details).

While B12 use different distance estimators, most of their statistics are based on a revision of the C10 distances, which we will call B12 distances and use throughout this work. In the following section we describe our calibration efforts. We would have preferred to use their distances, but our requests to obtain their adopted distances were declined. Therefore, to check the correctness of our derivation, we compare our result to the older data set of C10, to which we have access.

\begin{figure}
\begin{center}
\epsfig{file=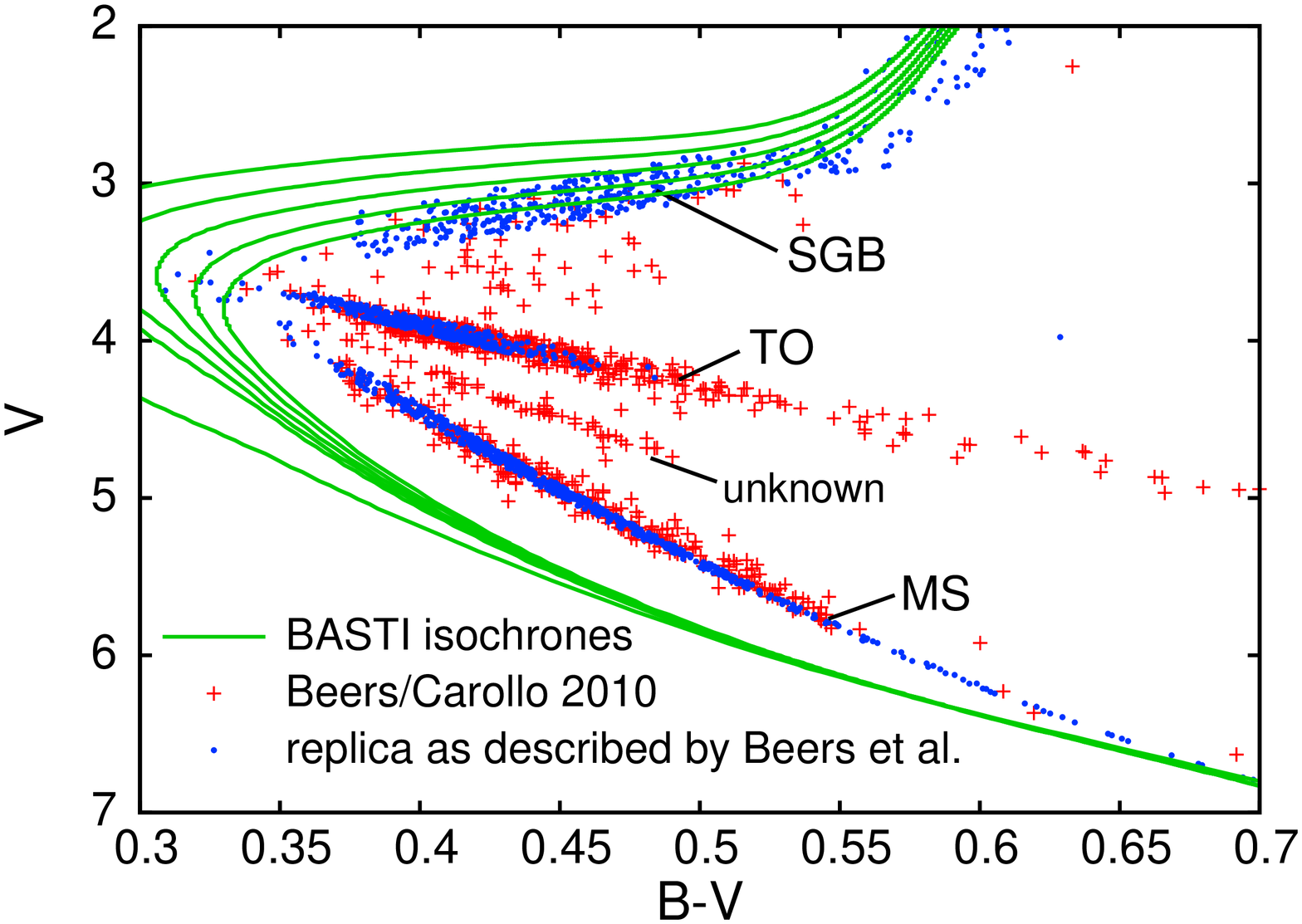,angle=-90,width=\hsize}
\epsfig{file=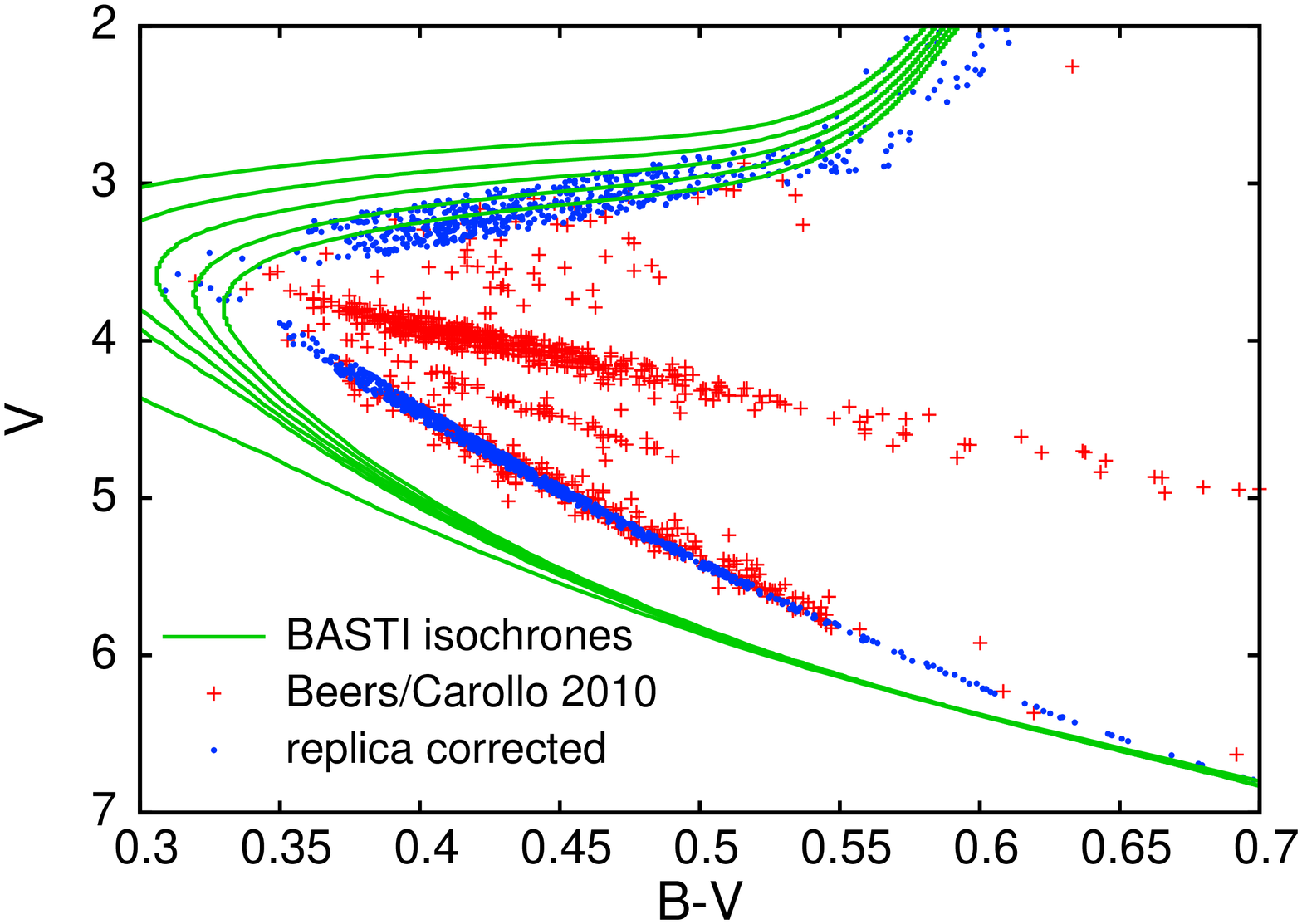,angle=-90,width=\hsize}
\caption{Reproduction of the B12 distance calibrations. The red points show the stars from C10, limiting $-2.4 < \feh_{C10} < -2.1$ on their own metallicity determination $\feh_{C10}$. We use their given SDSS colours for transformation into the Johnson-Cousins system, where we calculate their absolute magnitudes from the given distances. With blue circles we show our reproduction of the new B12 prescriptions, using data from SEGUE DR8. For comparison we plot BASTI isochrones at $1,9,10,11,12,13 \Gyr$ and metallicity $\feh =-2.27$ with green lines. In the top panel we replicate the prescriptions of B12, while the bottom panel displays the replicated data after correcting the critical temperature (see text).}
\label{fig:isccomp}
\end{center}
\end{figure}

\subsection{Distance method and replication issues}

The basic idea of the C07, C10 and B12 distance calibrations is the strategy of \cite{Beers00} to classify stars by their gravities and then sort them into different branches (luminosity classes). B12 still assign stars with $\llg > 4.0$ into the main sequence branch and stars with $\llg < 3.5$ into a subgiant branch. The intermediate objects with $3.5 < \llg < 4.0$ are considered to be ''turn-off`` stars placed on a sequence half-way in magnitude between the dwarf and subgiant branches. For the clearly unphysical objects redwards of the oldest possible turn-off point, B12 modified their approach by sorting stars with $3.5 < \llg < 3.75$ to the subgiant branch and objects with $3.75 < \llg < 4.0$ into the dwarf sequence. 

This re-sorting, which affects nearly all metal-poor turnoff stars, does not resolve the general problem of assigning a wrong luminosity class to a fraction of the objects. In the unphysical position intermediate between the true relations, they all had an error of $\sim 1 \magn$. Now they are in permitted locations, but their intermediate surface gravity determinations imply errors of $\sim 2 \magn$ (in both directions) for an unknown number of stars that are misidentified as dwarfs or subgiants. We know that there is a substantial number of misassignments both from the calibration on globular clusters \citep[i.e. errors in derived $\llg$ in][]{Lee08a, Lee08b} and from the distance statistics of \cite{SBA}.

The absolute magnitudes on the three branches are taken from the calibrations of \cite{Beers00}, which are, however, formulated in Johnson bands $B$ and $V$. Hence they transform the SDSS colours into Johnson colours via the linear relationships derived by \cite{Zhao06}. These neglect any non-linear terms like those found by \cite{Ivz07} and also any metallicity dependence, which are difficult to detect on the \cite{Zhao06} sample of only $58$ stars. 

\cite{Beers00} have only a coarse grid in metallicity for their colour-magnitude relations, which comprises just three different metallicity values. How they perform interpolation/extrapolation between these three points in metallicity is not described by \cite{Beers00} nor by B12, but from a comment in \cite{Ivz08} we conclude that the most likely scheme is cubic splines with the natural boundary condition. 

As we can not access the B12 distances, we cannot directly test our replica. However, we can compare to their older data, which are shown with red crosses in Fig. ~\ref{fig:isccomp}. For this plot we selected metal-poor objects with $-2.4 < \feh < -2.1$. All colours are transformed via the \cite{Zhao06} formula into the Johnson-Cousins system. The absolute $V$-band magnitudes are taken from the old C10 data. Also shown in Fig. ~\ref{fig:isccomp} are BASTI isochrones \citep[][]{Piet04, Piet06} for ages of $1, 5, 10$ and $13 \Gyr$ and $\feh = -2.27$. Comparison with the isochrones shows again the un-physical "turn-off" objects from C07 and C10 as well as the surprising fourth sequence (labelled "unknown" in Fig. ~\ref{fig:isccomp}) between dwarfs and turn-off stars. We cannot judge if the latter is addressed in their new calibration. 

Comparison to the published plots in B12 and the C10 data shows that the derived sequences (in particular the main sequence) are in the right place. However, the replicated data show a remaining stump of the turn-off sequence, which appears absent in Fig. 2 from B12. B12 define a critical temperature $T_{\rm crit} = T_{\rm TO} - 250K $, $250K$ colder than the temperature of the isochrone turn-off, below which they re-sort stars on the turn-off branch into the two other sequences (see equation 1 and Fig. 1 in B12). The only possible solution to the discrepancy appears to be {\it adding} the $250K$ instead of subtracting them. This gives the blue points in the bottom panel of Fig. ~\ref{fig:isccomp}. We will adopt this solution throughout the paper.\footnote{We could not obtain a clarification from the authors. We can, however, point out that fortunately the adopted solution is not central to the presented results.} 

Apart from the well-agreeing sequence locations, both the old C10 sample (red points in Fig. ~\ref{fig:isccomp} and Fig. 2 of B12 show significantly more magnitude scatter than our derivation. We cannot explain this scatter by metallicity differences, which should be small at such low metallicities. This is not a systematic shift, but likely enhances the Lutz-Kelker bias in their sample.

\begin{figure}
\epsfig{file=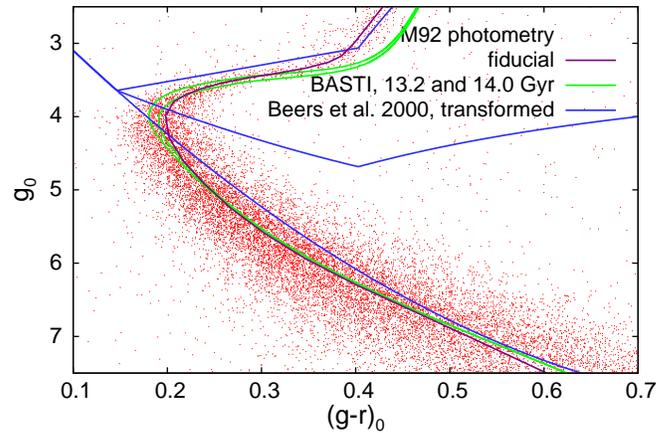,angle=-90,width=\hsize}
\caption{Comparison between BASTI isochrones with $Z=0.0015$ ($[Fe/H] = -2.1$) and age $13.2$ and $14 \Gyr$ (green lines), the subgiant, turnoff and main sequence from the Beers et al. (2012) calibration (blue lines) and the An et al. (2008) fiducials for the cluster $M92$. The cluster and its fiducial have been shifted using a distance modulus of $\mu_0 = 14.6$ and a reddening $E(B-V) = 0.023$ using the colour dependence of reddening from An et al. (2008).}
\label{fig:iscompz}
\end{figure}

\subsection{Origin of the main sequence discrepancies}

Reproducing the B12/C10 absolute magnitudes finally reveals the origin of the difference between C10 and the isochrones in SDSS colours: the \cite{Beers00} calibrations are brighter than the stellar models in Johnson colours: both the red points and the blue main sequence lie clearly above the isochrones in Fig. ~\ref{fig:isccomp}. Which one is correct? Though B12 state that the \cite{Beers00} distance calibration is based on Hipparcos parallaxes, this is only true for metal-rich stars. For metal-poor stars there are hardly any reliable Hipparcos parallaxes and none for globular clusters. Yet, there would be indirect calibrations for metal-poor clusters by using the few Hipparcos parallaxes for very metal-poor stars. This was done, e.g., by \cite{Pont98}, who derived a distance modulus of $\mu_0 = 14.61 \pm 0.08$ for $M92$, compared to a $\mu_0 \sim 14.57$ derived via isochrone fitting from \cite{Harris97}. Instead of using these Hipparcos derived parallaxes for their metal poor clusters, \cite{Beers00} relied on a calibration of horizontal branch star magnitudes. Since they did not report their derived distance moduli, we can not determine why their magnitude assignments deviate so far from well-calibrated models.

The \cite{Zhao06} transformation from Johnson to SDSS is significantly bluer (by about $0.1 \magn$ at low metallicities) than the predictions from model atmospheres entering the BASTI \citep[][]{Piet04, Piet06} models. This follows from its missing metallicity dependence, which forces a compromise between the different line-blanketing of metal-poor and metal-rich stars. The blue bias of the transformation for metal-poor stars helps to reduce the difference between the isochrones and the B00/B12 main sequence calibration in the SDSS system to about $0.3 \magn$, smaller than in the Johnson-Cousins system to which the B00/B12 calibration is native. Still we can see from Fig. ~\ref{fig:iscompz} that in contrast to the B00/B12 calibration the isochrones match the photometry for the cluster $M92$ very well on the main sequence with adopted standard reddening of $0.023$ and a distance modulus of $\mu_0 = 14.6$, in concordance with the Hipparcos-based calibration of $M92$. 

In summary, the bright bias of the B12 main sequence calibration in SDSS colours traces back to the calibrations of \cite{Beers00} being up to $0.5 \magn$ brighter than the isochrones in Johnson colours, mitigated to some extent by the transformations between the colour systems.

\section{Analysing the main arguments for a dual halo}\label{sec:arguments}

After replicating the B12 distance assignments, we can analyse their main three arguments for a dual Galactic halo: i) declining metallicity in their sample towards larger altitudes, ii) the decomposition of the observed azimuthal velocity distribution into Gaussians and iii) their argument that the proper motions of retrograde stars are highly different from the rest of the sample. In the following subsections we address each of these claims, arguing that none bears closer scrutiny.

\subsection{Trend of metallicity with altitude}\label{sec:radmet}

As laid out in the introduction, radial abundance trends are not a signature of halo duality and can be either caused by the specific accretion history \citep[][]{Cooper10} or by kick-up of disc stars. As the Sun is at a significant distance from the Galactic centre, such a radial gradient should in principle be reflected as well by a local gradient in altitude $z$. 

B12 claim that the metallicity of halo stars decreases with altitude $z$ (see their Fig. 13 and corresponding discussion) down to a peak metallicity of $\feh \sim -2.2$ at an altitude of $z > 9 \kpc$. They identify this as a clear transition from an inner to an outer halo in their picture of a halo duality. Bias control in such an investigation is key: not only do we expect altitude-dependent disc contamination; this sample exemplifies metallicity-dependent luminosity biases and selection effects. Here we will follow the standard route of statistical analysis: We derive bias controlled statistics in the next subsection and in the following parts provide the causal reasoning/backgrounds by analysing the exact origin of the bias affecting the B12 findings.

\begin{table}
\begin{tabular}{l|cc|cc} 
\multicolumn{5}{c} {\bf Linear fit parameters} \\
parameter & all & $\sigma_{\rm all}$ & $\feh < -1.4$ & $\sigma_{\feh < -1.4}$ \\\hline
$d\feh/dz$ & $-0.1350$ & $0.0019$ & $-0.0338$ & $0.0015$ \\
$\feh(z=0)$ & $-0.9389$ & $0.0054$ & $-1.7120$ & $0.0054$ \\ \hline
$d\feh/dz$ & $-0.1657$ & $0.0051$ & $-0.0039$ & $0.0048$ \\
$d\feh/ds$ & $0.0290$& $0.0045$ & $-0.0272$ & $0.0041$ \\
$\feh(z=0)$ & $-0.9555$ & $0.0060$ & $-1.7002$ & $0.0057$ \\
\end{tabular}
\caption{\rm Linear fits to the declining metallicity with altitude both for the entire sample and restricted to likely halo objects with $\feh < -1.4$. In the upper half we show results for the simple fit, while below we control for a distance dependence according to equation \eqref{eq:cfit}.}\label{tab:zfitpar}
\end{table}

\subsubsection{Simple statistical analysis}

\begin{figure}
\epsfig{file=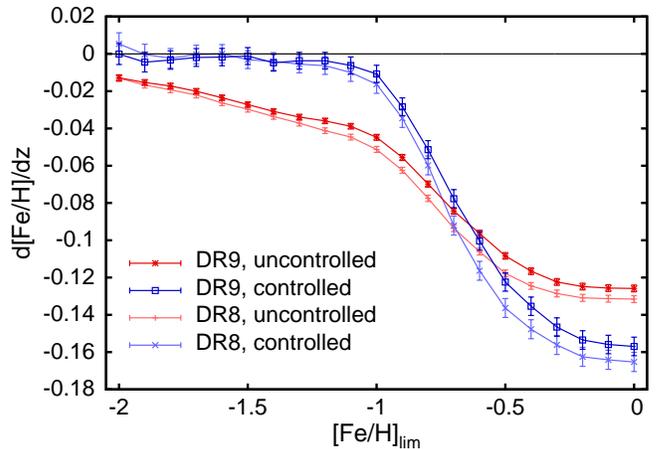,angle=-90,width=\hsize}
\caption{Controlled and uncontrolled (for the distance term) vertical metallicity trend in the sample out to $8 \kpc$. At each point statistics are taken exclusively for stars with $\feh < \feh_{\rm lim}$.}
\label{fig:ztrendlim}
\end{figure}

In the upper half of Table \ref{tab:zfitpar} we see that under a linear fit we recover a strong decrease of metallicity with altitude in agreement with B12. The linear fit may not reflect a perfect assumption, but gives a direct measure of the correlation between altitude $z$ and metallicities. However, most of this decline is caused by the transition from disc to halo towards higher altitudes. To isolate the Galactic halo, we limit our sample to $\feh < -1.4$, where the disc contamination at lower altitudes should vanish in the present sample as evident in Fig. 6 of \cite{S12}. We can see this also from the red line in Fig. ~\ref{fig:ztrendlim} - the curve has a kink at $\feh \approx -1$ where the bulk of disc contamination ends. The weakening trend towards even lower metallicities is caused less by disc contamination than the subsequent narrowing of the available metallicity range. At those low metallicities there is still a negative slope with a striking significance of $20 \sigma$. Is it real? We will show that this is not the case.

\begin{figure}
\epsfig{file=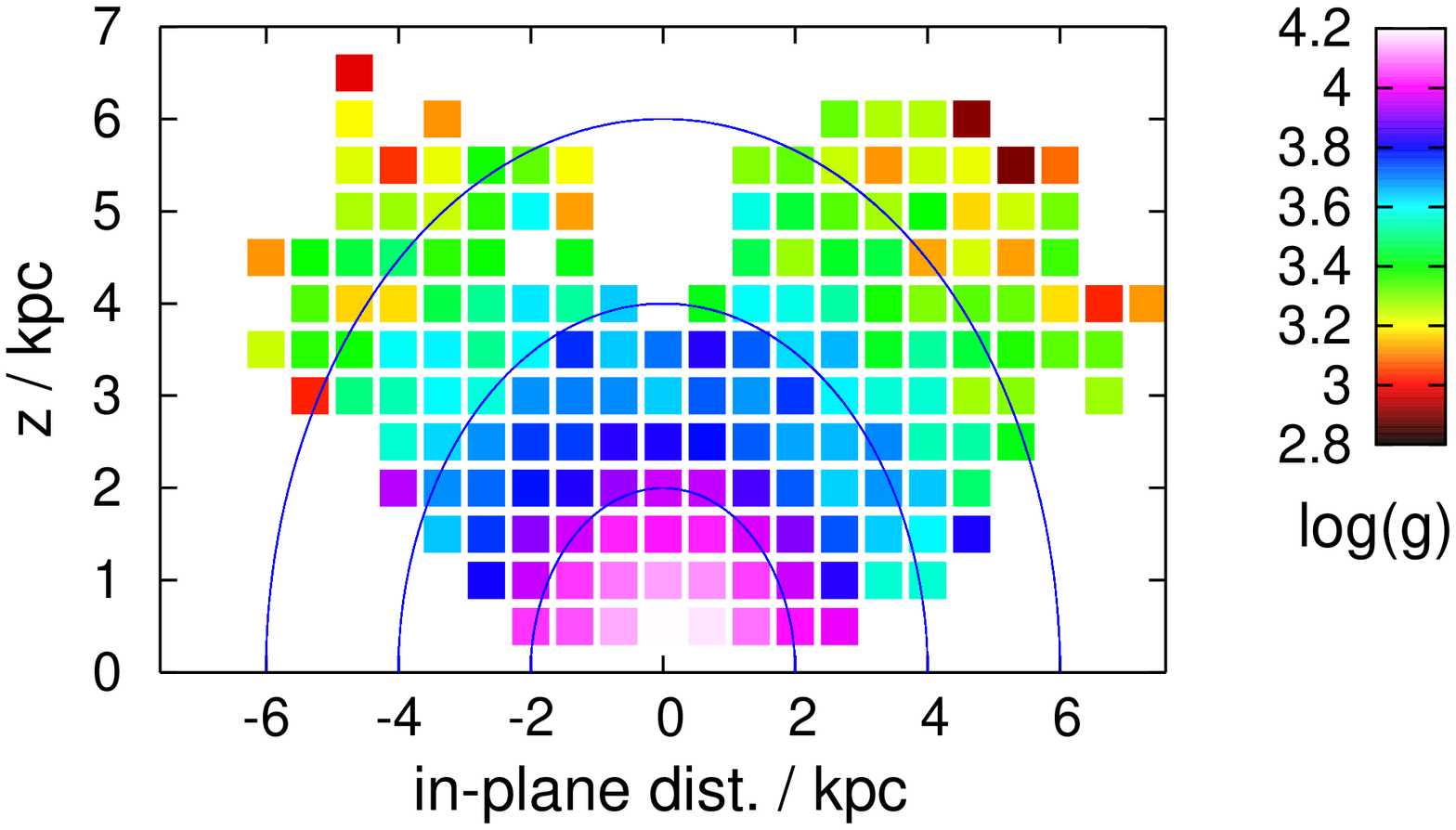,angle=-90,width=\hsize}
\epsfig{file=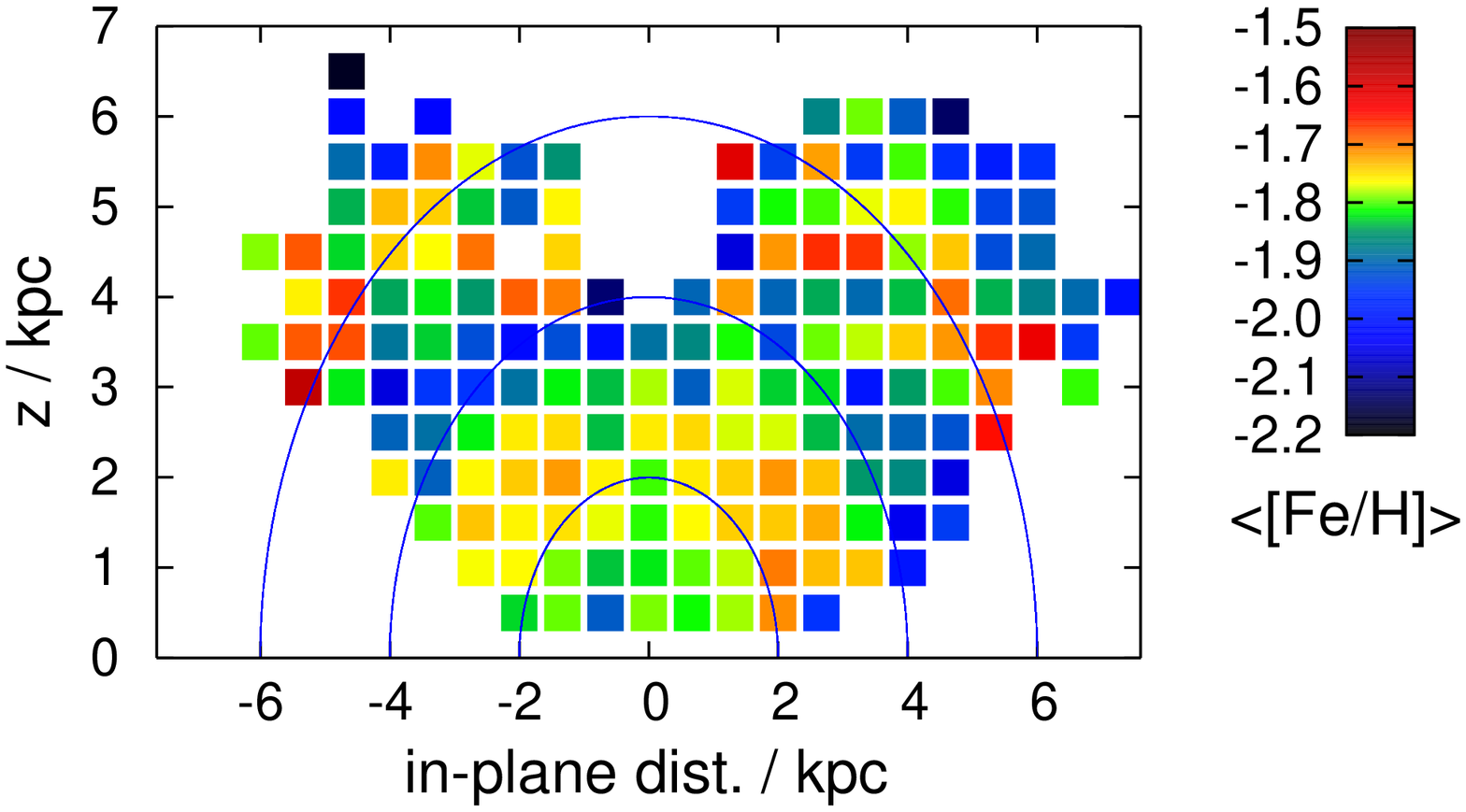,angle=-90,width=\hsize}
\caption{Upper panel: Binned mean gravities in a plane defined by the altitude $z$ on the y-axis and the in-plane distance $\sqrt{x^2 + y^2}$ on the x-axis, where stars inside the Solar galactocentric radius, i.e. with $R < \Rsun$ are assigned negative values and stars with $R > \Rsun$ positive values. The lower panel shows the mean metallicities. The blue lines depict equidistant circles of $s = 2, 4, 6 \kpc$. Only bins with more than $7$ stars were plotted.}
\label{fig:planestats}
\end{figure}

In statistics a well-known phenomenon is the omitted variable bias: Let there be a causal connection between a quantity $X$ and a quantity $Y$. However, we measure the dependence of $X$ on another variable $Z$, which is not causally connected to $X$, but correlated with $Y$. In this case we will infer an artificial correlation between $Z$ and $X$ after neglecting $Y$. The standard solution to this problem is controlling for the omitted variable by including it into a simultaneous fit of $X$ against $Y$ and $Z$. An omitted variable bias is demonstrated if the trend gets significantly altered or even vanishes after inclusion of the additional variable. Note however, that the reverse case is not true: Finding a trend with inclusion of one omitted variable does not prove its reliability, since there might be missing variables, or the wrong functional form is used in the fit.

In our case the three quantities are distance $s$, altitude $|z|$, and metallicity $\feh$. In any pencil-beam study centred on the Sun, the average altitude $|z|$ will by the geometry rise almost linearly with distance $s$, and hence the two quantities are strongly correlated. The strong polewards bias (i.e. most stars are at high Galactic latitudes $b$) of the SEGUE sample exacerbates this correlation. At the same time, $\feh$ is correlated with distance: stars with lower metallicity are brighter than more metal-rich stars of similar mass and age, so that metal-poor stars can be seen further away. In a general sample selection this competes with the colour shift: metal-rich stars are redder, so that at fixed colour metal-rich main sequence stars are brighter than their metal-poor counterparts. For the subgiant branch we only have the first effect, since metal-poor subgiant stars are in general brighter than metal-rich ones. In addition, the gravities of stars, as determined by the SEGUE pipeline, are significantly lower on the metal-poor end (for $\feh < -1 \dex$ the "main sequence" density ridge slopes by $\sim 0.4 \dex$ in gravity per $\dex$ in metallicity), and the observational error is larger (see Section \label{sec:ddd} below), making them more prone to be selected for the turn-off or subgiant branches in the B12 scheme with its fixed gravity cuts. 

Thus, in order to claim a trend of metallicities with altitude one must prove that this trend does not come from a selection effect in distance $s$. A simple solution is expanding the fit equation:
\begin{equation}
\feh_i = (d\feh/dz)\cdot z_i + \epsilon_i
\end{equation}
to
\begin{equation}\label{eq:cfit}
\feh_i = (d\feh/dz) \cdot z_i + (d\feh/ds) \cdot s_i + \epsilon_i $,$
\end{equation}
where $i$ is the index running over the stellar sample, $(d\feh/dz)$ and $(d\feh/ds)$ are the free fit parameters, measuring the correlation between metallicity and altitude $z$ as well as distance $s$ respectively, and $\epsilon_i$ are the individual deviations to be minimized.

The statistics for this test are shown in the lower half of Table \ref{tab:zfitpar} and are visualized in Fig. ~\ref{fig:ztrendlim}. Along the x-axis we vary an upper metallicity limit, selecting only stars with $\feh < \feh_{\rm lim}$. Indeed there is a strong trend of metallicity with distance. Simultaneously, the physical trend of metallicity with altitude when including disc stars is enhanced (blue error bars), while the altitude trend $d\feh/dz$ for halo stars vanishes. 

The bias can also directly be seen in Fig. ~\ref{fig:planestats}, where we bin the sample sample in altitude $z$ and in-plane distance $(x^2 + y^2)^{0.5}$. To account for a possible radial metallicity gradient, stars with $R < \Rsun$, i.e. inside the Solar Galactocentric radius, are given a negative sign. As apparent from the upper panel of Fig. ~\ref{fig:planestats}, the stars with lower gravities dominate the longer distance range as they are sorted into the turn-off or subgiant sequences (with limits $\llg = 4.0, 3.75$, and $3.5$). Since the $DR8$ pipeline has a strong correlation between metallicity and gravity (see Fig. 3 in SAC11), it comes as no surprise that we measured a strong distance bias in metallicity. This is evident in the lower panel of Fig. ~\ref{fig:planestats}, where we plot the mean metallicity of stars with pure halo metallicities, i.e. $\feh < -1.4$. Notice that instead of a vertical trend, the lower metallicity bins populate concentric rings around the Sun, associated with the boundaries of turn-off and subgiant star domination in the sample. 

\begin{figure}
\epsfig{file=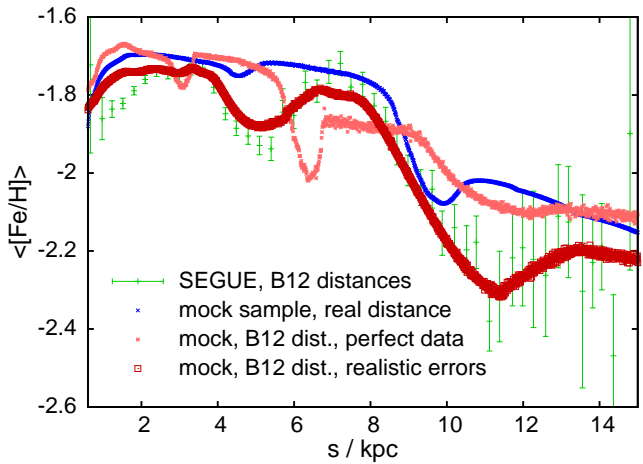,angle=-90,width=\hsize}
\caption{Mean metallicity of stars with $\feh < -1.4$ in the B12 sample depending on heliocentric distance with green error bars. For comparison we plot a measurement with a SEGUE-like colour and magnitude selection on a mock sample of a simple Galactic halo with a spatially constant metallicity distribution. Blue points show the mock measurement with accurate distances, red points the measurement using the B12 distances, where the dark red curve displays realistic errors, while faint red depicts the case of perfect metallicity and gravity measurements.}
\label{fig:bmetreplica}
\end{figure}

\subsubsection{Detailed distance dependence}\label{sec:ddd}

For a thorough analysis of the found distance bias we plot the profile of the mean metallicity versus distance in the DR9 sample with green error bars in Fig. ~\ref{fig:bmetreplica}. The metallicity profile is highly unphysical a priori and is proof by itself for the distance bias. It is, however, useful to nail down the origin of this odd structure. 

The local minimum just outside $4 \kpc$ that we saw in Fig. ~\ref{fig:planestats} divides two relatively metal-rich peaks, and beyond $8 \kpc$ the metallicity drops to a value of $\feh \sim -2.2$. To trace the origin of this bias, we use a mock sample: We assume a halo with a simple $1/R^3$ density profile and a spatially constant metallicity distribution to populate the colour-magnitude space with the $13 \Gyr$ BASTI isochrones\footnote{We use the isochrones at $Z=0.00015$, $0.0002$, $0.0003$, $0.00045$, $0.0006$, $0.00075$, $0.001$, $0.00145$, $0.002$, $0.0029$, $0.004$, $0.0057$ and populate them with a probability distribution of $P(\feh) \sim \exp(-(\feh-1.2)^2/(2\cdot0.55^2))$} out to $40 \kpc$. We also assume a constant reddening of $0.15 {\rm mag}$ in the $g$-band. This is not meant to be a fully realistic representation of the Galactic halo and the survey, but suffices for demonstration purposes.

On this sample we mimic the SEGUE observations. To match their colour distribution we select only stars with $0.12 < (g-r)_0 < 0.55$. The blue bias of the sample is modelled by multiplying with $\exp(-((g-r)_0-0.38)/0.045)$ redwards of $(g-r)_0 = 0.38$. The magnitude distribution of SEGUE is dominated by the division into faint ($g_0 \gtrsim 17$) and bright ($g_0 \lesssim 17$) observing plates. For our metal-poor subsample the observed magnitude distribution $\rho_g(g_0)$ is very well described by:
\begin{equation}
\hspace{-0.7cm}
\rho_g(g_0) = \left\{
\begin{matrix}
64.2(g_0 - 14.4)^2 - 1) && $for $ 15.4 < g_0 \le 17 \\
10.6(g_0 - 14.4)^2 - 1) && $for $ 17 < g_0 \le 18.5 \\
0 && otherwise
\end{matrix}
  \right.
\end{equation}

We bin the virtually observed sample in magnitude and weight each bin to achieve the observed magnitude distribution. The resulting metallicity profile is shown with a green line. This green line, which purely shows the selection bias of the sample, captures already the features of the observed data. Now we have to replicate the actual observations with their actual errors. We compare the DR9 results to gravity determinations from \cite{AllendeP08}, fitting a linear trend for the gravity difference depending on metallicity:
\begin{equation}
\Delta_{\llg} = \delta_0 + \delta_{\feh} \cdot \feh $,$
\end{equation}
where $\Delta_{\llg} = \llg_{\rm SSPP} - \llg_{\rm AP}$. The resulting fit parameters are $\delta_{feh} = 0.10 \pm -0.05$ and $\delta_0 = (-0.24 \pm -0.07)\dex$, i.e. the SEGUE DR9 results have $-0.25 \dex$ lower gravity at solar metallicity than \cite{AllendeP08}, and this difference widens by $0.1 \dex$ in gravity per $\dex$ lower metallicity. Then we "correct" the SEGUE DR9 data by this metallicity dependent offset and measure the trend in the absolute values of the residuals. This gives a slope of $-0.16 \pm 0.03$ dex in gravity per dex in metallicity, with a zero point offset of $(0.14 \pm 0.04)\dex$, i.e. the dispersion at solar metallicity is about $0.16 \dex$, widening to about $0.3 \dex$ at $\feh \sim -1$.

\begin{figure}
\epsfig{file=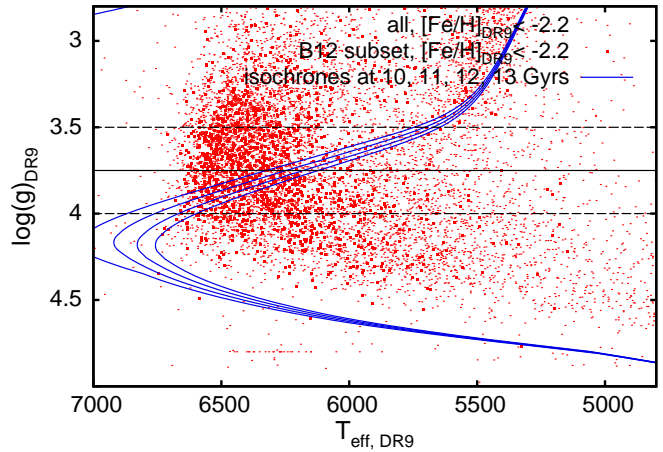,angle=-90,width=\hsize}
\caption{Gravities versus effective temperature from SEGUE DR9 for stars with $\feh < -2.2$. Red squares show the selected sample, red dots depict stars in the full SEGUE. In comparison to the BASTI solar scaled isochrones of ages $(10, 11, 12, 13 Gyrs)$ at metallicity of $[M/H] = -2.2$ (blue lines) the SEGUE gravities are too low by $\sim 0.5 \dex$ along the main sequence, a clean selection into sequences (the B12 gravity limits at $\llg = 3.5, 3.75, 4.0 \dex$ are shown with black lines) is made impossible by the large scatter.}\label{fig:fehloggsc}
\end{figure}

While selection effects bar us from a direct assessment of the gravity errors depending on metallicity near the turn-off, the trend in mean metallicity for the main sequence can be clearly identified in Fig. ~\ref{fig:fehloggsc}, where we plot the temperature-gravity plane of extremely metal-poor stars from the entire SEGUE sample as selected in \cite{S12} with red dots. Red squares highlight the stars ending up in the B12 sample selection. The hot bias of the latter is mostly caused by the blue bias of the photometric and reddening calibration star selection. In comparison to stellar model expectations (BASTI isochrones depicted with blue lines), the main sequence and subgiant branch gravities are too low by about half a dex, in good agreement with the findings from comparing to \cite{AllendeP08}. Though the main sequence is recognizable, a confident separation of stars into sequences with the criteria of B12 (black lines) is impossible.
In the following we will use $\delta_{\llg} = -0.1 + 0.2\feh$ and $\sigma_{\llg} = 0.15 - 0.12\feh$ as a rough estimate. The resulting metallicity profile is plotted in Fig.~\ref{fig:bmetreplica} with dark red.

The general structure is the same both with perfect distances and with the replicated B12 sample. The underlying metallicity distribution is constant in space, but the measurement induces strong Malmquist-like biases. The red sample cut-off leads to a metallicity rise at the smallest distances: Metal-poor stars on the main sequence are, at fixed colour, less luminous than their metal-rich counterparts and can thus enter the sample at smaller distances. Just beyond $4 \kpc$ the metallicity drops. At this point the turn-off, which essentially gives the maximum luminosity of the bulk of stars, reaches the faint cut-off of the sample. Since metal-poor populations have a brighter turn-off and subgiant branch, they will dominate the far part of the sample. The $4 \kpc$ drop is connected to the first magnitude limit of $g_0 < 17$ in the considered sample. Further out, the fainter sample selection dominates and gives rise to a second metallicity peak, followed by a drop beyond $8 \kpc$, where the sample only contains subgiants and giants. The effect is again exacerbated by the red cut-off, which further reduces the relative number of metal-rich stars in the sample. Compared to perfect distances (blue line) the B12 distances together with the pipeline errors (dark red line) enhance the metallicity drop around $s = 8 \kpc$.

While the dark red line approximates the qualitative structure of the observations remarkably well, there are mild quantitative differences. In particular the metallicity of the nearest stars is overestimated. This is certainly not a "real" trend, but derives from our rough assumptions like a fixed reddening, or the rough shape of the metallicity distribution. Indeed the exact shape of the replica depends quite strongly on the assumed pipeline errors, age structure, initial mass function and much more. Had we used "perfect" gravities and metallicities on our mock sample, the overall structure would be shifted massively inwards as depicted by the light red line in Fig.~\ref{fig:bmetreplica}: the gravity cuts of the B12 distances are deep in the subgiant regime and hence on clean gravity determinations would lead to a prevalence of distance underestimates. Any further refinement on the mock catalogue would hence primarily serve to understand the SEGUE errors and selection rather than advancing the discussion on halo structure.

Two important points should be taken away from this Section:
\begin{itemize}
\item{In a sample that is not volume complete, the metallicity distribution can not be explored without extensive bias correction. This is apparent from the fact that the unbiased sample (green line) already showed significant metallicity effects. Even if we had perfect distances, we would suffer from both the magnitude and colour selection and hence a spatially dependent bias would remain.}

\item{Our understanding of the detailed distance dependence of metallicities in the sample complements the statistical analysis from the previous section. It confirms that the data are consistent with no change in mean metallicity throughout the extent of the sample out to an altitude of about $z = 10 - 15 \kpc$. However, the statistical analysis provides a more robust value, since it filters the distance dependence without requiring knowledge of the detailed errors and biases in the sample. It shows that at $\feh < -1.4 \dex$ the data are consistent with a constant metallicity. Any vertical trend in these data is less than a third of the B12 claim, i.e. smaller than $\sim 0.01 \dex \kpc^{-1}$.}
\end{itemize}

\begin{figure}
\epsfig{file=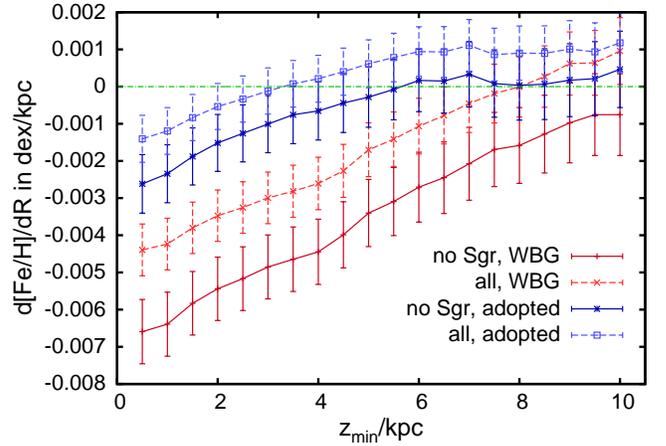,angle=-90,width=\hsize}
\caption{Radial metallicity gradient in the BHB sample (in dex/kpc) with varying cut for the minimum altitude $z_{\rm min}$ to exclude disc contamination. The red lines with error bars depict the Wilhelm, Beers \& Gray (1999) metallicities, while blue lines use the pipeline adopted values. Dashed lines show the same sample with inclusion of the Sgr region.}
\label{fig:xuebhbmet}
\end{figure}

\subsubsection{BHB star metallicity trend with Galactocentric radius}

B12 make a similar claim for BHB star metallicities, now not as a function of altitude but as a function of Galactocentric distance. The effect is very small (less than $-0.01 \dex \kpc^{-1}$, see Fig. ~\ref{fig:xuebhbmet}), and it is nearly impossible to reliably disentangle it from a distance bias. A clean separation further out in the halo is not reliable, because distance from the Sun becomes degenerate with the distance from the Galactic centre. Additionally the signal to noise ratio will affect the observed width of the metallicity distribution and related parameters. \cite{WBG99} report not only a larger error in measured line-widths at lower signal to noise ratios, but also a significant systematic shift due to issues with continuum placement. Hence it is reasonable to expect some metallicity shift with signal to noise ratio and hence distance. The \cite{WBG99} method is also vulnerable to small colour errors as small as $0.03 \magn$, which are expected at large distances (high observed magnitude) and at low latitudes (reddening uncertainty), reporting a $\sigma_{\feh}$ of up to $0.72 \dex$. Furthermore, distances to BHB stars depend significantly on their metallicity with metal-poor stars being about $\sim 0.3 \magn$ brighter \cite[][]{Piet04, FSI} in the g-band, such that the precautions of the above subsection fully apply.

We can also test for another influence on the trend: the Galactic disc. B12 cut in Galactocentric radius, while for their BHB sample they did not discuss and examine the dependence on the altitude above the Galactic plane. Especially the lower latitudes of the sample will carry a disc contamination that evokes a spurious trend. In addition it is not known how far contamination by the bulge extends into the Galactic halo, and the sample contains a significant number of stars at low galactocentric distances and relatively low altitudes ($R \sim 5 \kpc$, $|z| \le 6 \kpc$). Note also that the disc contamination can exceed the normal geometric limits of the disc because of a significant contamination with blue stragglers that have distance overestimates \citep[][]{Sirko04}. In our determination of the radial metallicity trend in the sample we hence vary the lower limit of the altitude $|z|$ as plotted in Fig. ~\ref{fig:xuebhbmet}. Indeed, for the \cite{WBG99} metallicities we encounter a very mild negative trend of metallicity with Galactocentric radius $R$ for the entire sample. However, the trend diminishes successively while we move up the cut to censor regions of likely disc contamination. Further, the trend noted by B12 only exists when using the \cite{WBG99} metallicities. If we use instead the pipeline adopted metallicities from SEGUE DR8 (blue lines in Fig. ~\ref{fig:xuebhbmet}), the trend starts at a very small value and becomes already insignificant at a limiting altitude of $z_{\rm lim} \sim 3 \kpc$. We note, however, that there is an increase in the metallicity spread noticed by an increase in the number of both objects with $\feh < -2.5$ and $\feh > -1.5$ at larger distances. The dispersion increase might be real, but we cannot cleanly disentangle it from a natural increase in the observational error.

In summary, we cannot detect any reliable downtrend of mean metallicity with Galactocentric radius that is robust against a disc contamination.

\subsection{The azimuthal velocity distribution of halo stars}\label{sec:VD}

The observed azimuthal velocity distribution is frequently used to decompose an observed data set into single components. SAC11 and \cite{SB12} emphasized the fact that a typical disc velocity distribution is not symmetric and has a long tail towards lower velocities, providing new and physically motivated fitting functions to replace the unphysical Gaussian functions.
 
B12 tried to circumvent this problem by examining the velocity distribution of stars they can by metallicity classify as halo stars and again attempt a Gaussian decomposition. However, this does not address the fundamental failure of decompositions in terms of Gaussian distributions to characterize inherently non-Gaussian distributions, which was one of our main concerns raised in SAC11.
\begin{itemize}
\item There is expectation that the halo azimuthal velocity distribution should be Gaussian. Dynamic friction may influence the shape, as well as the relaxation time for single halo stars is longer than a Hubble time, so a fully thermalized halo velocity distribution cannot be expected.
\item The observational error distribution that has to be folded with the underlying velocity distributions to obtain the final observed velocity distribution is highly non-Gaussian. While the non-Gaussianity caused by using photometric parallaxes has been known at least since \cite{Strom27}, it is commonly known as a \cite{Lutz73} bias, who demonstrated its existence for geometric parallaxes. It is inevitable that the B12 sample is affected by this bias and correction for it is essential, which however was not done by B12 even though it was one of the key concerns we had articulated in SAC11.  
\end{itemize}

The non-Gaussian error on distances translates directly into a non-Gaussian deformation of the velocity distribution. The long tail of distance overestimates translates to a long tail of the velocity distribution away from the solar velocity, i.e. into the retrograde regime. This was already pointed out by \cite{Strom27} and more recently by \cite{Ryan92}. This distortion is not only caused by distance overestimates: distance underestimates, and misidentifications of evolved stars as dwarfs yield a distribution skewed in the same way, as they create a narrower shifted onto the prograde side. An attempt to fit such a left-skewed distribution at moderate data precision with Gaussians will usually produce an unphysical split into a strong component centred around the maximum and a weaker component out in the long tail.

In addition, proper motion errors themselves are highly non-Gaussian (though usually symmetric) due to a major magnitude dependence of astrometric precision \citep[][]{Munn04, Dong11}. These proper motion errors multiply with the varying distance and angle terms, so that velocity errors generated by them are highly non-Gaussian.

\begin{figure}
\epsfig{file=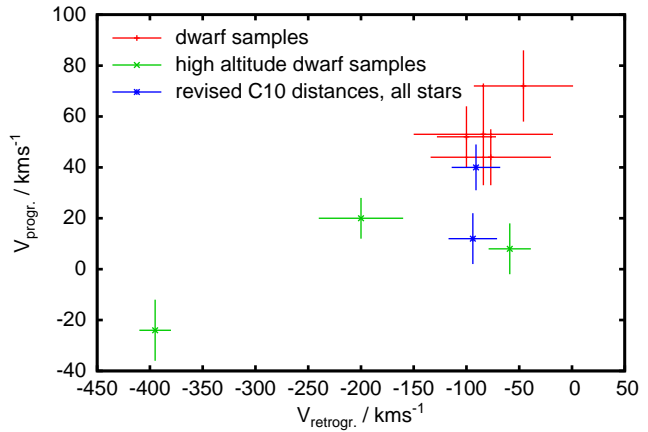,angle=-90,width=\hsize}
\caption{Plot of the mean velocities of the two components in the Gaussian fits by B12. They identify their more retrograde component with their "outer", more retrograde halo and the prograde component with its "inner" counterpart. The larger number of data points results from the different distance calibrations they use, their "revised C10" sample is their distance calibration discussed in this paper.}
\label{fig:Beersfits}
\end{figure}

A typical outcome of trying to fit data with such unphysical functional components is large scatter in parameters and unphysical parameters \citep[cf.][]{Johnson92} that cannot be properly explained by just the moderate distance shifts between the different methods. This is exactly what can be observed in the results of C10 and B12. Fig. ~\ref{fig:Beersfits} shows the mean azimuthal velocities of their low (i.e. more retrograde and identified as an outer halo) component on the x-axis against the mean azimuthal velocities of their high component (more prograde, their inner halo component) on the y-axis under their different selections and applied distance scales. The red error bars are their general dwarf samples, while the green error bars restrict those samples to stars that in their analysis reach altitudes in excess of $5 \kpc$. Unsurprisingly, the latter samples favour more retrograde components, because they have a bias towards stars that are further away and have a high proper motion impact on $V_\phi$ (and have hence lower accuracy in their kinematics), as well as stars with distance overestimates. Blue error bars depict the full sample with their new distance calibration. Clearly the exact rotational velocities for the two claimed halo components differ wildly depending on the chosen distance prescriptions and how exactly the analysis of C07, C10 and B12 is carried out.

\begin{figure}
\begin{center}
\epsfig{file=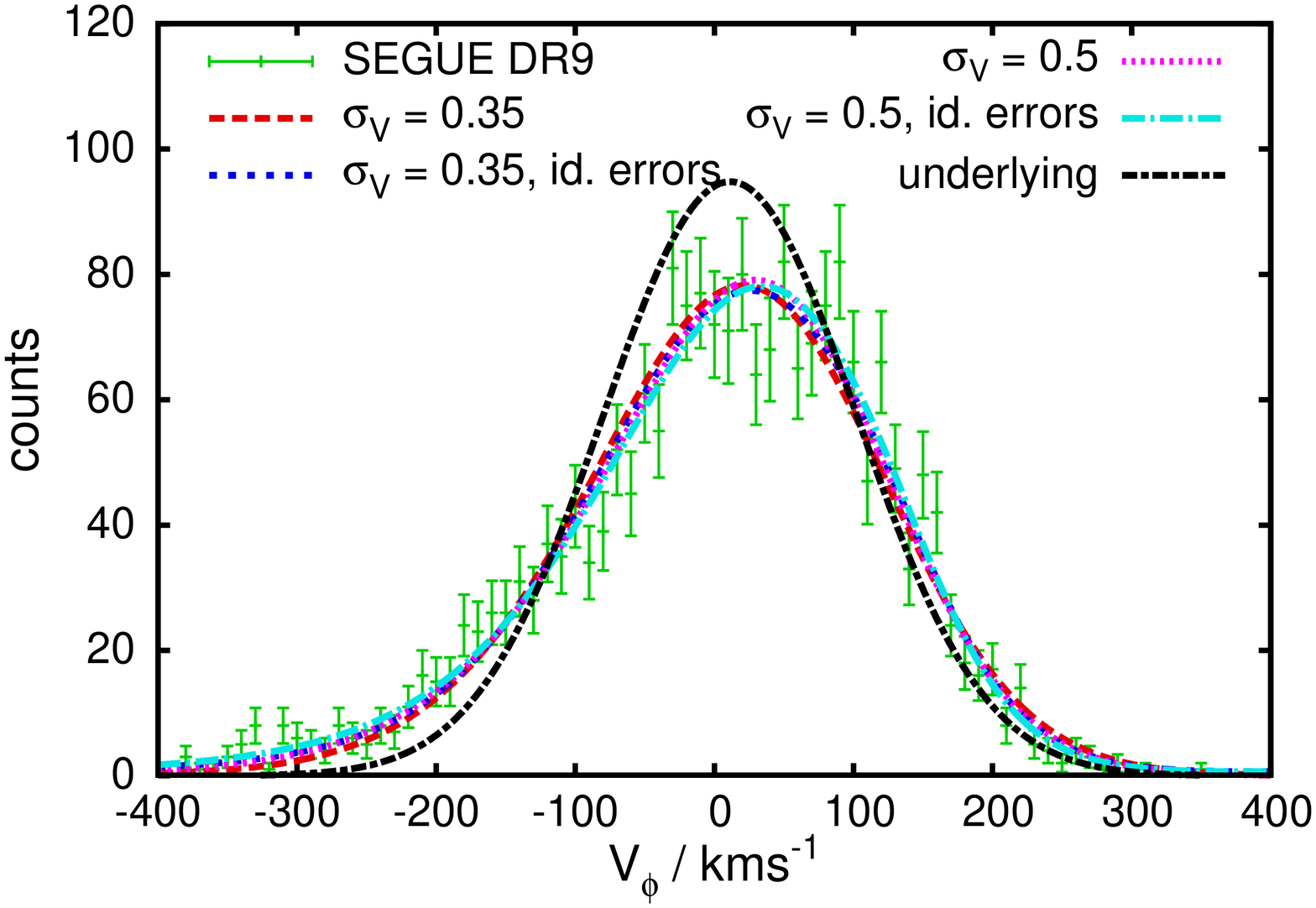,angle=-90,width=\hsize}
\epsfig{file=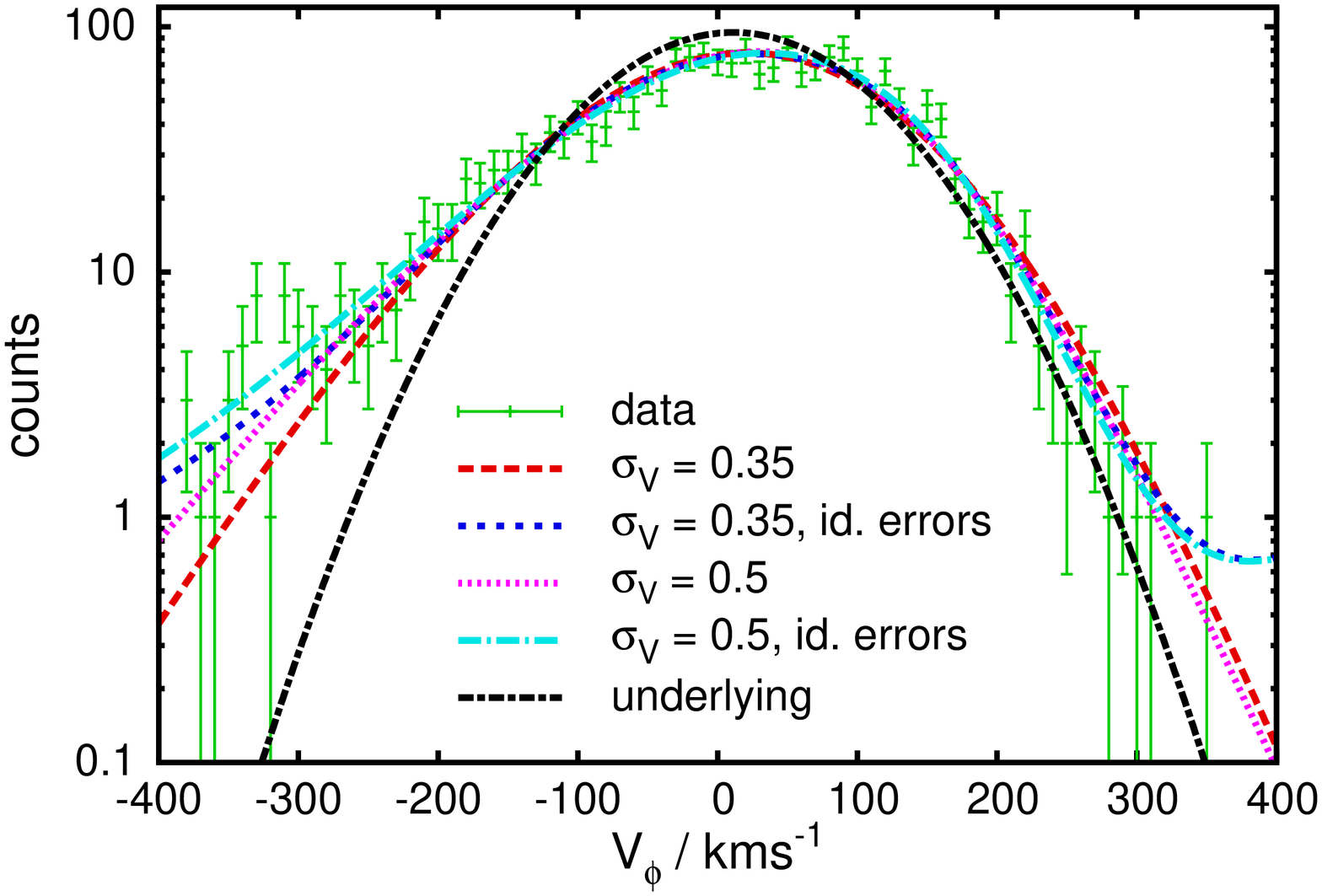,angle=-90,width=\hsize}
\caption{Azimuthal velocity distribution in the subsample with $\feha < -1.9$. We show simple fits with two different values for the magnitude uncertainty $\sigma_{V}$, as well as cases (marked as ''id. errors``) where we assume magnitude overestimates and underestimates of $1.2 \magn$ in addition comprising each $5 \%$ of the sample. For the last model we also show the underlying distribution (not folded with the error term).}
\label{fig:Vanothertime}
\end{center}
\end{figure}

To assess the halo velocity distribution with an appropriate error analysis, we make the crude assumption of an underlying Gaussian distribution and fold it with the following error terms:
\begin{itemize}
\item The proper motion errors, which can be treated as approximately Gaussian for single stars. We calculate the distribution of velocity errors resulting from the positions and proper motion errors of each single star and fold it onto the velocity distribution.
\item A correction for distance errors. Those are highly non-Gaussian if we assume a Gaussian magnitude error. We use values of $\sigma_{V} = 0.35 \magn$ and $\sigma_{V} = 0.5 \magn$, corresponding to a distance error of approximately $17\%$ and $26 \%$.
\item A smaller correction for velocity crossovers via the distance uncertainty as discussed by \cite{SBA}, which should be approximately symmetric in velocity space. We keep this at a conservative $7 \kms$.
\end{itemize}

Fig. ~\ref{fig:Vanothertime} shows the azimuthal velocity distributions of stars with metallicities of $\feh < -1.9$. For our fits we use the mean velocity, the width, and the normalization of the underlying Gaussian term as free parameters. For both values of $\sigma_{V}$ we include also a case where we model the identification errors that will occur in the sample. For this case we use both an underestimated and an overestimated fraction, each comprising $5 \%$ of the total sample and having an additional $1.2 \magn$ offset. This is a rather conservative estimate, regarding the fact that a main-sequence-subgiant identification error amounts to $\sim 1-3 \magn$ (cf. Fig. ~\ref{fig:isccomp}). Similarly, the number of misidentified stars is chosen far on the low side. Most objects are near the turn-off and very hot and metal-poor. For these stars, gravity determinations are very difficult, due to the disappearance of suitable temperature and pressure broadened wings at the typical signal to noise ratio and spectral resolution for SEGUE. This renders the correct identification of luminosity classes very difficult. This is also shown by their distance statistics. The linear distance estimator from \cite{SBA} flags an average $35\%$ distance overestimate\footnote{Note that simply $5\%$ contamination with $100\%$ distance overestimates gives an average $5\%$ distance overestimate.} at $7\sigma$ significance for the subgiants with $\feh < -1.9$, opposing a fair average distance for the declared dwarfs. For the latter, the misidentified subgiants balance the overestimated main sequence distances.

We show the underlying velocity distribution fit for the latter model. The differences are subtle, but a higher $\sigma_V$ expands the left wing while shifting the peak a little bit into the prograde regime, whereas the identification errors mostly show up at the broader wings. The comparison to the unfolded distribution (black) demonstrates how much more the left (retrograde) wing is enhanced by the symmetric magnitude errors. Neglecting the observational error asymmetry would drive at least one other (artificial) component in a Gaussian decomposition, as indeed we argue has happened in C07, C10 and B12. 

The determinations for the underlying parameters of the velocity distribution are dominated by systematic errors: The dispersion values vary between $98 \kms$ for $\sigma_V = 0.35 \magn$ without identification errors and $91 \kms$ for $\sigma_V = 0.5 \magn$ accounting for identification errors. Since we likely underestimate the uncertainties the real azimuthal velocity dispersion of the halo should be around $90 \kms$ or markedly lower. The mean azimuthal velocity varies between $11 \kms$ and $14 \kms$, consistent with a non-rotating or at best weakly prograde halo. However, this is a gross underestimate of the systematic uncertainty. The value shifts by $18.5 \kms$ when distances are shifted by $10 \%$, in addition to the uncertainty in the solar azimuthal velocity of another $8 \kms$. Note also that as a consequence of their stark distance overestimates the subgiants display a mean azimuthal velocity about $60 \kms$ lower than the dwarf stars at the same metallicity. These problems can be partly remedied by better quality cuts and, by applying the distance corrections of \cite{SBA} and by using robust estimators. However, this has been already done by \cite{S12} for the local sample and \cite{FSII} for BHB stars, who found a non-rotating or at best weakly prograde Galactic halo.

Within the expected magnitude and identification uncertainties, the wings of the distribution can be fully explained by just the observational errors, without having to deviate from the simplistic underlying Gaussian velocity distribution (which would not yet imply the presence of a second component). Hence the azimuthal velocity distribution offers not the least hint of any halo duality.

\subsection{Detection in proper motions} 
\begin{figure}
\epsfig{file=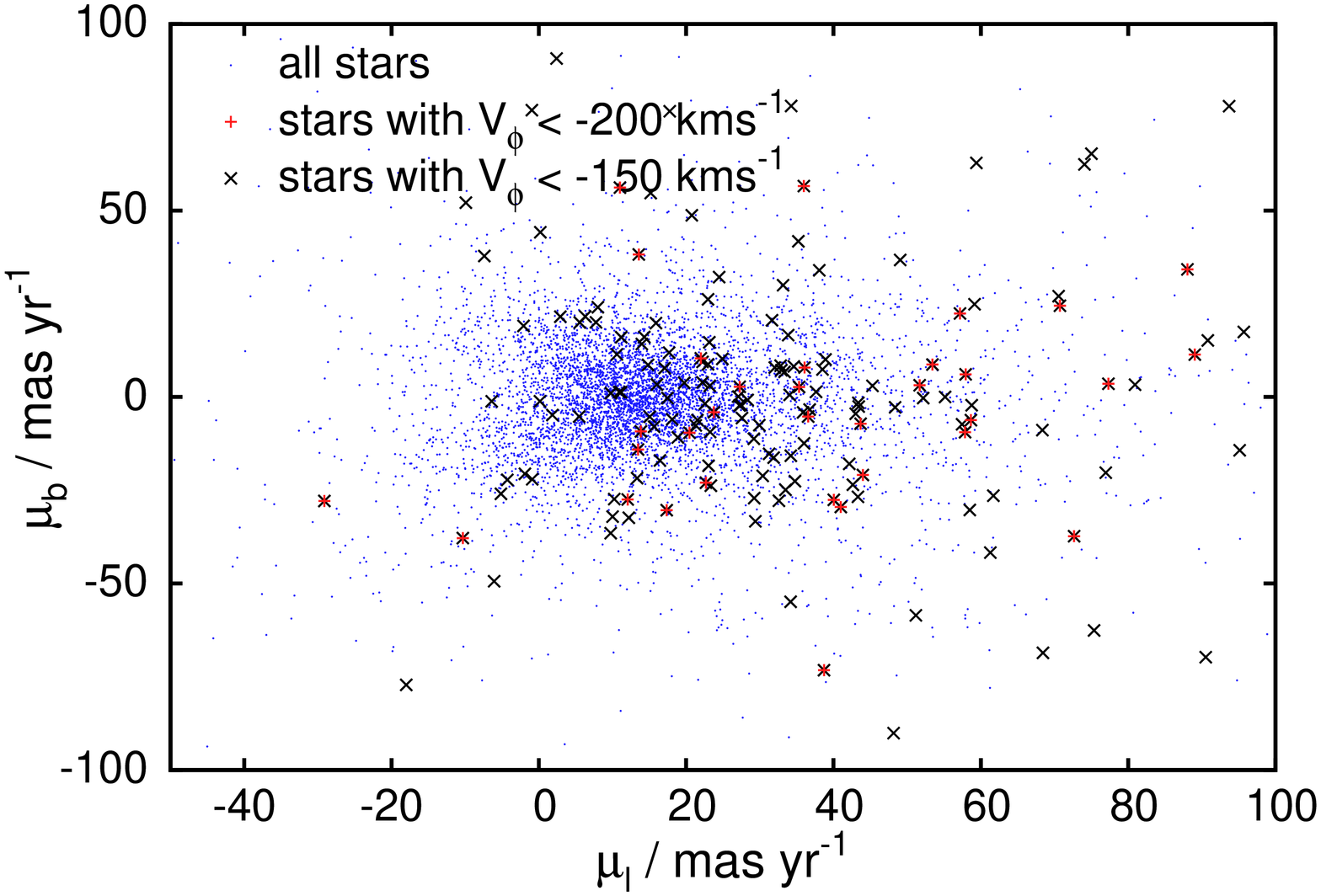,angle=-90,width=\hsize}
\epsfig{file=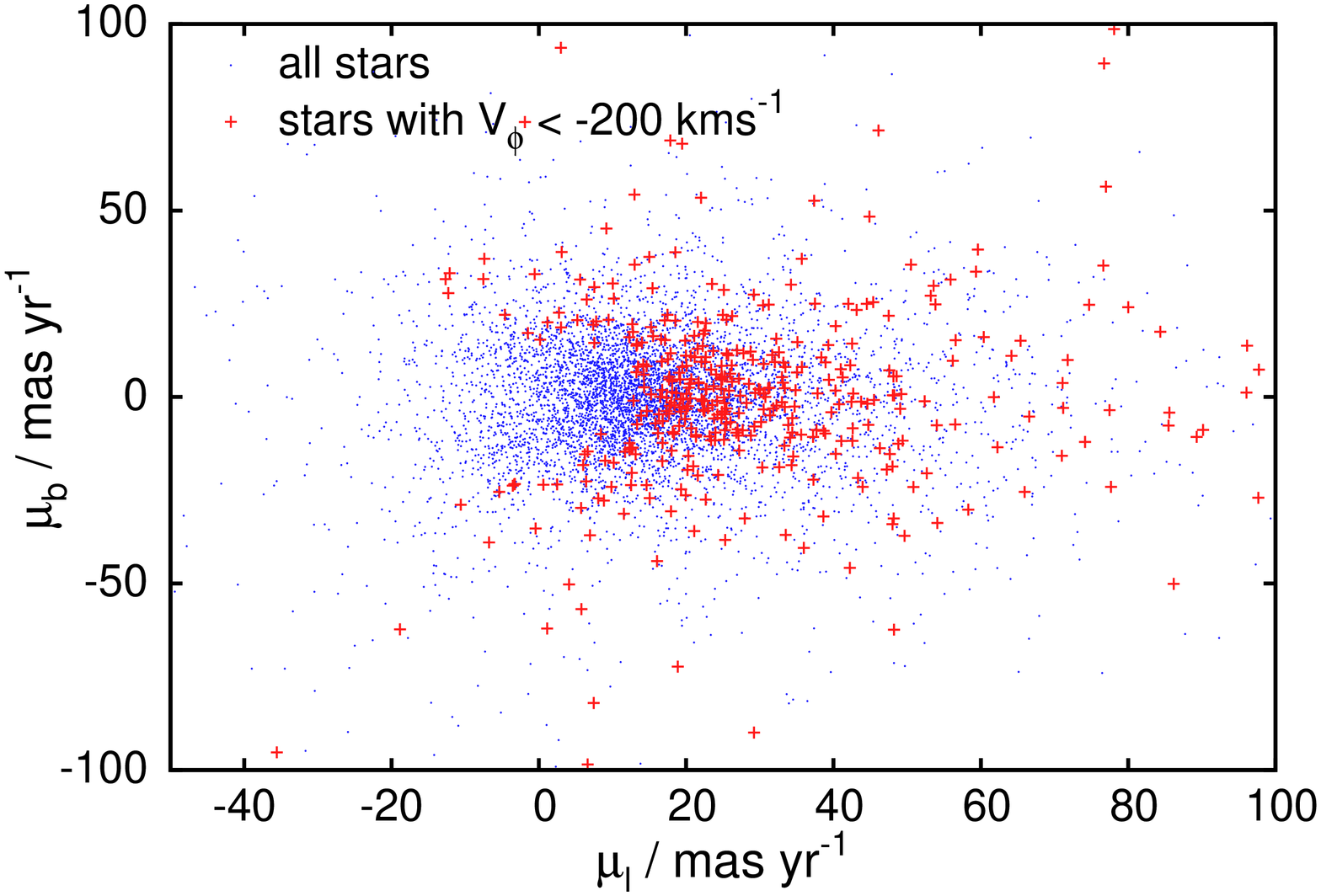,angle=-90,width=\hsize}
\epsfig{file=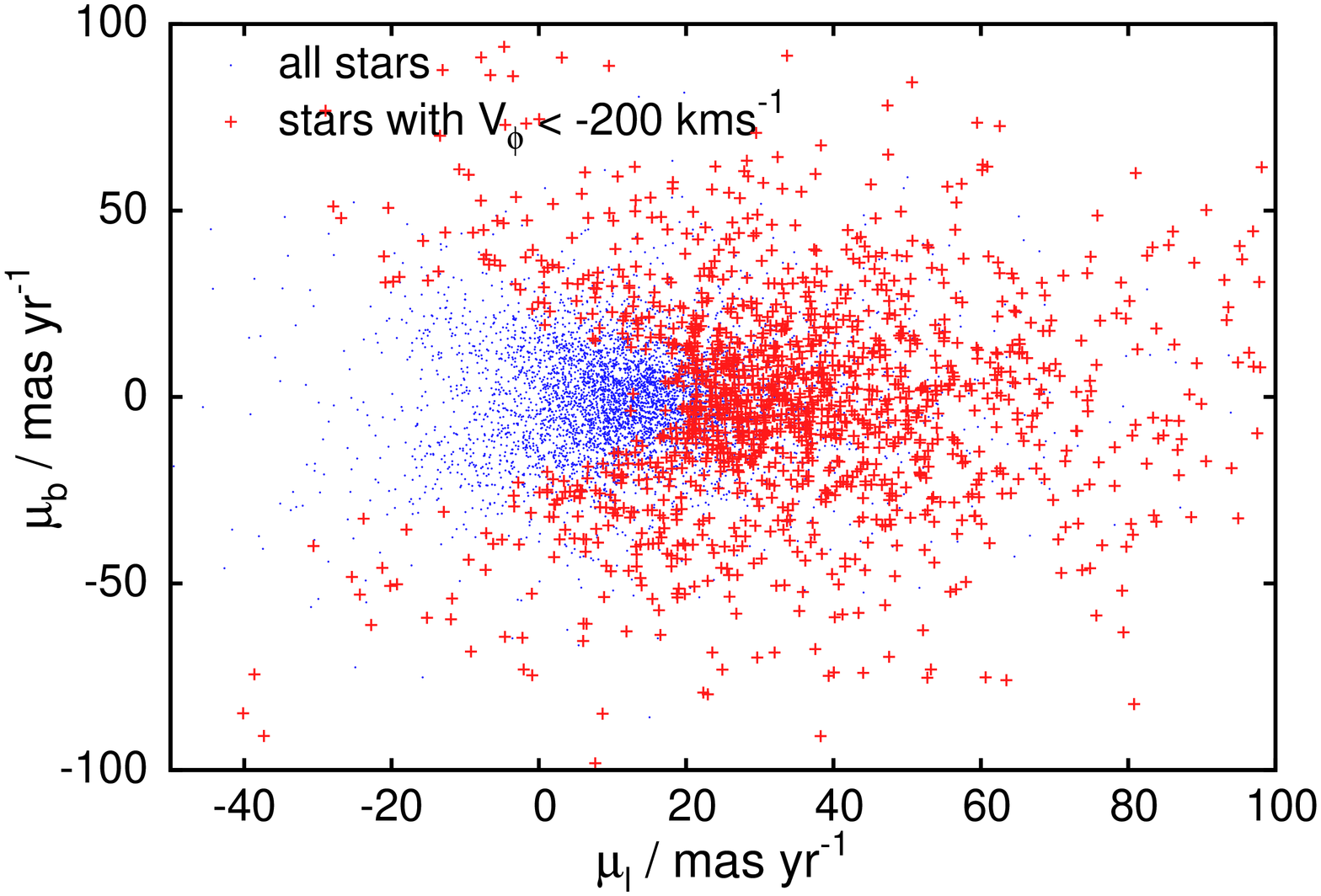,angle=-90,width=\hsize}
\caption{Proper motion plane in a mock sample. The top panel shows a perfect ''measurement`` of the mock data, in the middle panel we applied proper motion and radial velocity errors of $3.0 \masyr$ and $5 \kms$, a systematic distance error of $15\%$ and a Gaussian distance error of $25 \%$. To avoid negative terms, distance offsets are capped at $-90\%$. In the bottom panel we apply the observational errors and assign fully random distances with $0.7 \kpc < s < 5 \kpc$ to all stars. In all samples we highlight the highly retrograde stars with $\V_{\phi} < -200 \kms$. Since there are only $\sim 25$ real cases, we highlight in addition the stars with $\V_{\phi} < -150 \kms$ in the top panel.}
\label{fig:scrambdist}
\end{figure}

B12 argue that their retrograde halo stars are real, because their distribution in the proper motion plane is different from the rest of the sample. 

Specifically, B12 argued that {\it "if the retrograde signature were indeed created in the manner suggested by S10 alone (and the stars we assign to a highly retrograde tail were otherwise identical in their kinematic properties to the rest of the low-metallicity halo stars), we would a priori expect their observed proper motions to be drawn from the same parent population as those not in the tail."}

The strategy to refute this argument is to construct a mock sample and to show that distance errors produce many false retrograde objects, while these objects display a remarkable difference in proper motion space compared to the rest of the sample.

We note first that even with a single Gaussian-shaped halo velocity distribution centred at zero we would expect $0.5$ to $1\%$ of all stars to have $V_\phi < -200 \kms$. Hence we do not claim that all strongly retrograde stars are spurious, but that the majority of observations in this region is a consequence of observational errors. Further, it is easier to move a star by observational errors from $-100 \kms$ to $-200 \kms$ than from $50 \kms$. Hence the "a priori" claim of B12 that a Lutz-Kelker bias draws stars equally from the entire parent population is a priori wrong.

Distance errors actually tend to enhance the contrasts between the thought-to-be retrograde tail population and the parent population, especially in a sample like SEGUE, where the polewards bias makes proper motions the dominant part in the azimuthal velocity measurements. Distance errors in a measurement affect exclusively the part of motion relative to the Sun that is carried by the proper motions.\footnote{We can directly see this from Eqs. (5) and (6) in \cite{SBA}: 
\begin{equation}
\begin{pmatrix} U \cr V \cr W\end{pmatrix} =
 {\bf M} ({\bf I}+ f{\bf P}) \begin{pmatrix}s\mu_b \cr s\mu_l \cr v_{\parallel} \end{pmatrix},
\end{equation}
 where ${\bf I}$ is the identity matrix, $s$ the true distance, $f$ the fractional distance error and 
\begin{equation}
{\bf P} \equiv\begin{pmatrix}1&0&0\cr0&1&0\cr0&0&0\end{pmatrix}$.$
\end{equation}
If a star has negligible proper motion, its distance has nearly no impact on the derived velocity vector.
}
Thus, the presence of distance errors is only able to sweep stars into spurious retrograde orbits when these stars have a considerable proper motion, i.e. the likelihood of a star being wrongly sorted into retrograde motion vanishes at the origin (0,0) in the proper motion plane and generally increases with proper motion. In turn, the density contrast between stars with high and low $V_\phi$ around the origin of the proper motion plane will commonly intensify in the presence of distance errors.

Let us examine this effect in a mock sample depicted in Fig.~\ref{fig:scrambdist}. Following \cite{SBA} we use a sample of $25000$ halo stars with a Gaussian distribution with zero mean velocity and dispersions $(\sigma_U, \sigma_V, \sigma_W) = (150, 75, 75) \kms$. If we plotted the entire sample, it would be fully centred around the origin of the proper motion plane. To make the test resemble more the SEGUE sample we only select stars towards the Galactic anticentre ($\cos l < 0$) and in the Northern Galactic hemisphere ($b > 0$). Again this is not supposed to be a realistic Galaxy, nor a good match to SEGUE, which has a narrower and more ragged angle distribution.

The result, assuming a perfect measurement without any errors, is shown with blue dots in the top panel of Fig.~\ref{fig:scrambdist}. Red data points depict the strongly retrograde stars with $V_\phi < -200 \kms$. Since there are only $\sim 25$ such stars, black crosses highlight in addition the stars with $V_\phi < -150 \kms$ to visualize the spatial distribution. In the middle panel we apply typical errors for a sample, using $3 \masyr$ in both proper motion components, $5 \kms$ in the line-of-sight velocities, combined with a systematic distance error of $+15\%$ plus a random distance error of $25 \%$. As a consequence the number of stars deemed to be retrograde multiplies by a factor of $\sim 10$. For the reasons laid out above the spurious members are at high proper motion, while virtually no false detections get added at lower proper motions, where the proper motions are just too small to change the velocities considerably (as a back-of-the-envelope calculation, consider that $10 \masyr$ at $2 \kpc$ distance correspond to $\sim 100 \kms$ in transverse velocity, which have to be multiplied with the geometrical factor ($< 1$) and the fractional distance error). 

To consider a really extreme case, we assign completely random distances with $0.7 \kpc < s < 5 \kpc$ to each star in the bottom panel. Though virtually all of our detections are wrong, the contrast between the sample distributions is actually enhanced (the relative density around the origin in the retrograde sample has measurably diminished) with similar asymmetric distribution in the proper motion plane as the case without any observational errors but obviously with many more spurious retrograde objects and with more pronounced asymmetry in azimuthal velocities. Such an asymmetry was argued by B12 to clearly demonstrate that the highly retrograde stars must be drawn from a different parent distribution than the stars with less pronounced rotation. The above illustration plainly shows that no such conclusion can be inferred from their proper motions. In particular, one can not directly interpret structure in the proper motion plane together with kinematic cuts. 

The argument turns even in the opposite way: the prominent proper motion structure seen by B12 would, if anything, indicate that many of their highly retrograde objects are artificial.

To summarize, none of the arguments of B12 bears up under closer scrutiny. In none of these statistics we can find even a hint of a halo duality; instead, this discussion serves as a prime example for the importance of sound error analysis and the danger of hidden biases.

\section{Discussing further arguments for a dual halo}\label{sec:evidence}

\cite{Deason11} claimed a difference in rotation between more metal-poor and more metal-rich BHB stars. While it is not clear support for the very particular claims of C10 and B12 of a retrograde, metal-poor {\it outer} halo, it seemed to support a rotation difference between metal-poor and metal-rich BHB stars while not finding a radial dependence. \cite{FSII} re-analysed this work and found that this result vanishes when stricter quality criteria are applied to the data. They also found that different estimators of rotation are inconsistent on the original sample and no reliable difference in rotation can be detected. Similarly, \cite{FSII} found that there is no trend of rotational tendency with Galactocentric distance, which contrasts to the findings of \cite{Kinman12}. We suspect that the latter result can be ascribed to the very small sample size, the very narrow range in lines of sight, especially with no line-of-sight velocity support for $V_\phi$ measurements and the use of partially untested proper motions like SDSS DR7 \citep[][]{DR7}, which are now corrected in DR9 \citep[][]{DR9}.

Similarly, the notion of B12 about the Galactocentric radial velocities of BHB stars in the sample of \cite{Xue11} demands investigation (see their Fig. 14). While the B12 claim is not backed by a correct statistical test, this issue was examined by \cite{FSII}. There is actually a statistically significant difference within the sample, but it is not a clean distinction between the metal-poor and the more metal-rich BHB stars. It derives from a difference within the metal-poor population, i.e. between the hottest and the cooler metal-poor BHB stars, while the cooler BHB stars show no difference towards any metal-rich subsample. Hence this is either interesting substructure in the halo, or a problem with SEGUE line-of-sight velocities. At the spectral resolution and typical signal to noise ratios of SEGUE, hot and metal-poor BHB stars display no metal lines for the determination of their line-of-sight velocities, such that the pipeline has to rely on the very broad Balmer resulting in large uncertainties. Hence, the difference points to an unreliable spectral analysis. Furthermore \cite{FSII} examined the subsample in question and found no evidence for a retrograde motion when using more reliable 3D-estimators for rotation. 

\cite{Jong10} found a strong radial metallicity gradient in a photometric study of the Galactic halo. We note that such a gradient does not constitute halo duality. \cite{Jong10} did not correct selection effects and so without studying their sample in detail we face a similar selection bias as B12: metal-poor stars are brighter than their metal-rich counterparts, thus they can be observed further away. Their metallicity transition happens around $15 \kpc$ distance, where the distance modulus brings the turn-off magnitudes into the proximity of the faint cut-off and higher magnitude errors. Since their most metal-poor isochrone is nearly $1$ magnitude brighter than their intermediate isochrone, there will be an increasing impact of the lowest metallicity on their analysis, an effect they do not correct for. More importantly, we can directly see the pattern expected from a selection effect in their figures. In their Fig. 6 the metallicity transition between their two "halo populations" is not only roughly circular around the Sun in the in-plane-distance vs. altitude plots, but more importantly it happens near an in-plane-distance $D = 15 \kpc$ at all Galactic angles $l$. This translates into a Galactocentric Radius $R = 14 \kpc$ for $l = 70$ in contrast to $R = 23 \kpc$ for $l = 187$. To reconcile the two angles, the transition for $l = 70$ would have to take place around $D = 25 \kpc$ and hence outside their plots.
Without a rigorous demonstration that their trends is not a selection effect (and hence that we live in a rather particular halo that is strongly elongated along the Sun-Galactic centre connection while appearing roughly circular around the Sun), we speculate that we are facing the same phenomenon of a spurious trend with radius as in the B12 analysis and explained in Section \ref{sec:radmet}.

The increase of substructure with Galactocentric radius, as found by \cite{Schlaufmann10} is expected from many theoretical models: the material in the outskirts of the Galactic halo is expected to have been accreted on average later and experiences less friction/scatter by the Galactic disc and central regions so that naturally should display more substructure. Hence, the results of \cite{Bell08} and \cite{Schlaufmann10} are expected. However, that does not provide evidence specifically for a dual halo with a retrograde, metal-poor outer component.

Finally, we note that the presence of theoretical papers advocating the physical possibility of a halo duality is interesting, but by no means constitutes any evidence that this is the case for the Milky Way, especially since the theoretical community is not unanimous in this respect.

\section{Conclusions}\label{sec:conclude}

We have studied the arguments presented by B12 for a dual halo with a retrograde, metal-poor outer component and conclude that none of them stands up to critical examination.

In the course of our replication of the B12 analysis we found that the relations from \cite{Beers00} are significantly brighter than stellar models in the Johnson V-band. In addition we checked their performance on M92 and could not reconcile the Beers et al. distances with the available data. Hence we caution against use of the calibrations of \cite{Beers00} for metal-poor stars. 

We also point out that B12 wrongly ascribed to us and named after us a distance relation that is far shorter/fainter than anything we ever used, just to show that it is too faint. 

The trend of metallicity with altitude claimed by B12 is primarily driven by disc stars. The smaller trend for low metallicities that exclude those disc objects is identified as a selection bias: metal-poor stars are more luminous and in SEGUE are more likely identified as brighter subgiant and turn-off stars in this sample. Hence the metallicity has a spurious correlation with altitude via its dependence on distance. Controlling for this effect, the altitude dependence of halo star metallicities vanishes. We identified an analogous problem for BHB stars and found a likely contamination with disc objects.

The selection bias investigated in Section \ref{sec:ddd} is an excellent example of why forward modelling from theoretical models is needed and why measured distributions from observations in a non-volume complete sample are unreliable without a thorough analysis and correction of biases.

The difference of proper motion distributions between retrograde stars and the remaining sample is not an argument that their data must be reliable, since stars with zero proper motions require extremely high line-of-sight velocities to be retrograde (and a position near the Galactic azimuth): retrograde stars essentially must show large proper motions. In fact, the contrast between ''retrograde`` (as measured) stars and the remaining population sharpens in proper motion space with decreasing sample quality.

Furthermore, B12 used unphysical Gaussian analysis on the azimuthal velocity distribution. As emphasized in SAC11, such analysis is inappropriate as the observational error is highly non-Gaussian and has long tails, especially towards the retrograde side. We demonstrated - now on the highly metal-poor stars - that we can easily fit it by an underlying Gaussian distribution folded with the appropriate measurement errors. 

The B12 finding of a difference between the velocity distributions of BHB stars does not trace back to a different behaviour between metallicities as they argue, but to hot metal-poor BHB stars behaving differently from their less hot metal-poor counterparts. Further checks do not reveal a reliable retrograde signal.

Further evidence in the literature is either subject to similar statistical and selection biases or not specific to the claim of a dual halo. For example, neither does the presence of increased substructure in the outskirts of the Galactic halo (which is a natural consequence of accretion and dynamic friction) constitute an indication for a retrograde, metal-poor outer halo, nor does the theoretical possibility of a duality as suggested by some (but far from all) simulations act as any observational evidence. 

While we find no evidence for any chemokinematic halo duality in the SEGUE sample, this disproof of evidence should not be misinterpreted as a proof of absence. Even with bold extrapolation the sample does not allow investigations beyond a distance of roughly $15 \kpc$. Further data and investigations especially with more remote samples, like the BOSS \citep[Baryon Oscillation Spectroscopic Survey][]{Dawson13} spectra, should be gathered to search for more subtle trends or trends beyond the current distance limits.

We repeat our statement from SAC11 that we favour a halo that shows an increase of substructure towards larger Galactocentric radii and do not try to address the question if there is any metallicity gradient towards its outskirts beyond what is probed by SEGUE. What we can definitely say is that there is no evidence from current SEGUE data that would support the claims of C07, C10 or B12 of a dual halo with a retrograde, highly metal-poor outer halo.

\section{Acknowledgements}

It is a pleasure to thank James Binney and David Weinberg, as well as the referee for very helpful comments to the text. R.S. acknowledges financial support by NASA through Hubble Fellowship grant HF-$51291.01$ awarded by the Space Telescope Science Institute, which is operated by the Association of Universities for Research in Astronomy, Inc., for NASA, under contract NAS 5-26555. The research of M.A. has been supported by generous funding from the Australian Research Council in the form of a Laureate Fellowship (grant FL110100012).

\end{document}